\documentclass[11pt,letterpaper,english]{article}
\usepackage[latin9]{inputenc}
\usepackage{textcomp}
\usepackage{amsmath}
\usepackage{amssymb}
\usepackage{graphicx}
\makeatletter

\pdfpageheight\paperheight
\pdfpagewidth\paperwidth
\advance \topmargin by -\headheight
\advance \topmargin by -\headsep

\evensidemargin \oddsidemargin
\newcommand{\lyxmathsym}[1]{\ifmmode\begingroup\def\b@ld{bold}
  \text{\ifx\math@version\b@ld\bfseries\fi#1}\endgroup\else#1\fi}

\makeatother

\usepackage{babel}
\begin{document}
\title{\textbf{Analytical Solution for Turbulent Flow in Channel}}
\date{Date: July 10, 2023}

\maketitle
\begin{center}
Alex Fedoseyev\\
Ultra Quantum Inc., Huntsville, Alabama, USA\\
Email: af@ultraquantum.com
\end{center}
\begin{abstract}

In this work the exact and approximate analytical solution of the GHE for turbulent flow in channel are presented. It was discovered first by numerical simulations,  Fedoseyev and Alexeev (2010), and now the explicit formula are obtained. The solution is a superposition of the laminar (parabolic) and turbulent (superexponential) solutions. The analytical solution compares well with the experimental data by Van Doorne (2007) for axial velocity and data by Nikuradse (1933) for axial velocity, for flows in pipes. 

It is proposed to explain the nature of turbulence as oscillations between the laminar (parabolic) and turbulent (superexponential) solutions. Good comparison of the analytical formula, a difference of the parabolic and superexponential solutions, for turbulent velocity fluctuations with the experiment by Van Doorne (2007) confirmed this suggestion. The Navier-Stokes equations do not have the superexponential solution.

The obtained analytical solution provides a complete structure of the turbulent
boundary layer that compares well with the experiments by Wei and Willmarth (1989). It also presents an explicit verifiable proof that Alexeev{'}s generalized hydrodynamic theory (GHE) is in close  agreement with experiments for turbulent flows.
\end{abstract}

\section{Introduction\label{sec:intro}}

Generalized Hydrodynamic Equations (GHE) have been proposed by Boris Alexeev (1994)\cite{Alexeev_1994}.  They were used for the simulations of incompressible
viscous flows for a wide range of problems and flow parameters, including
high Reynolds number turbulent flows with thin boundary layers in
3D driven cavity flow at Re = 3200 and 10,000, 2D backward facing
step flow at Re = 132,000, flow in channels for Reynolds number up to $Re=10^6$, magnetohydrodyamic flows \cite{Fedoseyev_1998,Fedoseyev_2000,Fedoseyev_2001a,Fedoseyev_2001b, Fedoseyev_2012}, resulting in good agreement with experiments \cite{Koseff_1984}, \cite{Kim_1980} and other works. 

GHE model has been applied to compressible hypersonic flows that exhibit
both continuum and non-continuum flow regimes. The shock wave (bow
shock) can be detached from the vehicle at high altitude, and near
boundary slip-flow is typical for such regimes. Results for
hypersonic GHE model have been reported in \cite{Fedoseyev_2020, Fedoseyev_2021, Fedoseyev_2022} are in close 
agreement with the experiments by \cite{Allegre_1997, Harvey_2001,  Harvey_2003} and other works even for highly rarefied flows.

In this work an approximate analytical solution of GHE for turbulent flow in channel is presented and compared with the experimental data.

Generalized Hydrodynamic Equations (GHE) are based on a new set of conservation equations obtained from a generalized version of the Boltzmann equation (GBE) by Alexeev that includes more details of the molecular collision processes \cite{Alexeev_1994}, \cite{Alexeev_2004}. The model combines continuum-to-free molecular flow physics in one consistent formulation by accounting for the kinetic effects (intermediate Knudsen number, fluctuations and turbulence) in the continuum approximation. A brief outline of the basic ideas of GBE and derivation of governing Generalized Hydrodynamic Equations is presented in Section  \ref{sec:GBE}.
\subsection{\label{sec:GBE}Generalized Boltzmann Transport Equation (GBE)}

Physical derivation of the standard Navier-Stokes equations (NS) can be obtained from the kinetic theory of gases, which is based on the solution to the Boltzmann transport equation for space-time evolution of particle velocity distribution function, $f$, written in the form
 
\begin{eqnarray}
\label{eq:BE}
\frac{ Df}{Dt} = J,
\end{eqnarray}

 where $D/Dt$ represents material derivative in space, velocity space and time and $J$ is the collision integral. The standard Boltzmann transport equation takes into account the changes in distribution function $f$ on hydrodynamic and mean time between collision time scales of infinitesimal particles. Accounting for a third time scale associated with finite dimensions of interacting particles gives rise to an additional term in the Boltzmann transport equation resulting in a generalization of the form as

\begin{eqnarray}
\label{eq:GBE}
\frac{ Df}{Dt} - \frac{D}{Dt}(\tau^{*}\frac{ Df}{Dt})= J,
\end{eqnarray}

where $\tau^{*}$ is the mean time between particle collisions. The new term is thermodynamically consistent and is proportional to the Knudsen number, $Kn$, and therefore in the hydrodynamic limit, to viscosity. More details on the GBE are provided in the Alexeev's book \cite{Alexeev_2004}.

\section{\label{sec:GHE}Generalized Hydrodynamic Equations}

Hydrodynamic equations can be obtained from Eq. (\ref{eq:GBE}) by multiplying the latter by the standard collision invariants (mass, momentum, energy) and integrating the result in the velocity space.  These equation are for incompressible viscous flow, presented originally in \cite{Fedoseyev_2012}, are the following:

%
%

\begin{equation}
\it \frac{\partial \bf V} {\partial \rm t} + ({\bf V}\nabla) {\bf V}
- Re^{-1}\nabla^2 \bf V + \nabla \rm p - {\bf F} = 
\tau \left\{ 2 \frac{\partial}{\partial t}(\nabla \rm p) + 
\nabla^2 (\rm p \bf V) + \nabla(\nabla \cdot (\rm p \bf V)) 
\right\}
\label{momeq}
\end{equation}

\noindent
while continuity equation is

\begin{equation}
\it \nabla \cdot \bf V = 
\tau \left\{
2 \frac{\partial}{\partial t}(\nabla \cdot {\bf V})
+ \nabla \cdot ({\bf V} \nabla){\bf V}
+\nabla^2 \rm p -\nabla \cdot {\bf F}
\right\}
\label{newconteq}
\end{equation}

\noindent
where ${\bf V}$ and $p$ are nondimensional velocity and pressure, ${Re=V_0 L/\nu}$ - the Reynolds number, $V_0$ - velocity scale, $L$ - 
hydrodynamic length scale, $\nu$ - kinematic viscosity, ${\bf F}$ is
a body force and a nondimensional ${\tau = \tau^* L^{-1}V_0}$.
Note that the right-hand side of (3) is the divergence of the 
fluctuation part of flow velocity, expressed explicitly through original 
primitive variables according to \cite{Alexeev_1994}.

We made the following assumptions deriving Eq. (\ref{momeq}, \ref{newconteq}):
\begin{itemize}
\item $\tau$ is assumed to be constant,
\item Neglected the nonlinear terms of the third order in the
  fluctuations, and terms of the order $\tau$/Re and smaller,
\item  Assumed slow flow variation, so neglect second derivatives in time.
\end{itemize}

Additional boundary conditions on walls are for fluctuations to be zero.
The boundary condition for pressure on  walls is

\begin{equation}
 ({\nabla \rm p - {\bf F})\cdot {\bf n}= 0} ,
\end{equation}

where ${\bf n}$ is a wall normal.

Note, that the dimension of a product $\tau^* \nu$ is
a square of length. We introduce fluctuation length scale $l$,
$l^2 = \tau^* \nu$ and rewrite nondimensional ${\tau}$ as
${\tau  = l^2 L^{-2} Re = K \cdot Re}$, where ${K = l^2 / L^2}$.
The value of $\tau^*$ is a material property and not known in advance, but we provide speculation 
on a choice of $\tau^*$ value in Section \ref{sec:general} and in the discussion of results, Section \ref{sec:discussion}.

The obtained GHE model is not a turbulence model, and no additional 
equations are introduced. Kinetic effects (small flow scales) have been successfully captured with the GHE,
and the obtained small scale of turbulence compared well with observations
in experiments by Koseff and Street (1984) \cite{Koseff_1984}, and 2D and 3D Naver-Stokes
solutions and k-$\varepsilon$ turbulent model solutions have been outperformed
by GHE results, Fedoseyev and Alexeev (2012)\cite{Fedoseyev_2012}.

In this paper to obtain the analytical solution we further simplify the GHE equations: 
(a) temporal derivatives are neglected in the 
fluctuations, (b) the  nonlinear terms are neglected in the 
fluctuations.

\subsection{\label{sec:2D}Governing Equations for 2D Incompressible Flow}

The case of 2D incompressible fluid flow is considered. The equations are taken from \cite{Fedoseyev_2012} where they are presented explicitly, and further simplification is done by dropping all the terms (with coefficient $\tau$) in momentum equations, and keeping only Laplacian  term in the continuity equation.
 
The resulting continuity equation of GHE model is the following:
\begin{eqnarray}
u_{x}+v_{y}&=&\tau\nabla^{2}p \label{cont}
\end{eqnarray}

while the momentum equations are :

\begin{eqnarray}
u_{t}+u\,u_{x}+v\,u_{y}+p_{x}&=&Re^{-1}\nabla^{2}u \label{mom1}\\
v_{t}+u\,v_{x}+v\,v_{y}+p_{y}&=&Re^{-1}\nabla^{2}v \label{mom2}
\end{eqnarray}

where $Re=LV_{0}/\nu$  is Reynolds number, $L$ is the length scale, $V_0$ is  velocity scale, $\nu$ is kinematic viscosity,  $\tau=\tau^{*}V_{0}/L$ is the nondimensional time scale from 
GHE ($\tau^{*}$ is the dimensional time).

One may point out that the continuity Eq. (\ref{cont}) is not correct, and the mass is not conserved for incompressible fluid (may have sources and sinks). Let us note, that 
\begin{itemize}

\item Eq.(\ref{cont}), as well as Eq.(\ref{mom1}, \ref{mom2}) are obtained by a the standard procedure, by multiplying Eq. (\ref{eq:GBE}) by standard collision invariants (mass, momentum, energy) and integrating the result in the velocity space. Then the simplification was done, and only the highest derivative was left in Eq. (\ref{cont}).

\item  Recall that the particles now are not the material points, but finitely-sized particles, that may be partially in and partially out of any control volume, and that is the origin of GBE, and the pressure Laplacian term in Eq.(\ref{cont}).
 
\item  When solving Eq.(\ref{cont}, \ref{mom1}, \ref{mom2}) numerically \cite{Fedoseyev_2010}, the residuals of solution of Eq.(\ref{cont}) with Laplacian term and without that term have been verified. The residuals have been small and of the same order of magnitude in both cases, so no sources and sinks have been observed for wide range of flows. oth cases, so no sources and sinks have been observed for wide range of flows. Still the results using the Navier-Stokes equation with div({\bf V})=0 were far from the experimental data while the GHE results with Eq.(\ref{cont}) fit   the experimental data well \cite{Fedoseyev_2010}.
\end{itemize}

\section{Laminar Flow Solution}

The 2D channel flow problem: a stationary flow in $x$
direction in a 2D (horizontal) channel of width $L$ with a center line
velocity $V_{0}$ is considered. So the time derivatives are dropped in Eq.(\ref{mom1}, \ref{mom2}). 
We consider the flow where everything is the same in any $x$ cross section, so all the partial derivatives in x are zero, except for the applied pressure gradient $p_x$.
The  boundary conditions are: 

$u=0$, $v=0$ and the normal derivative of the pressure $p_n=0$ at the walls $y=0$, and $y=L/2$.

A flow with $p_{x}=$ const has a parabolic velocity $u$ profile:

\begin{eqnarray*}
p_{x}&=&Re^{-1}u_{yy},\\
u_{y}&=&Re \ p_{x}y+C_{1} \\
u&=&\frac{1}{2}Re \  p_{x}y^{2}+C_{1}y+C_{2}.
\end{eqnarray*}

\noindent
Taking into account the boundary conditions, the laminar flow solution is

\begin{eqnarray}
U_{L}=u(y)=4V_{o}y(L-y)/L^{2}, \label{poiseille}
\end{eqnarray}

\noindent
where $V_{0}=-\frac{1}{8}Re\ p_{x} L^2$, and $v=0$ everywhere.

\noindent
This solution of Eq.\eqref{cont},\eqref{mom1},\eqref{mom2} is
also a solution of the Navier-Stokes equations, as $\nabla^{2}p=0$.

\begin{figure}[ht!]
\begin{center}
\includegraphics[width=0.48\textwidth]{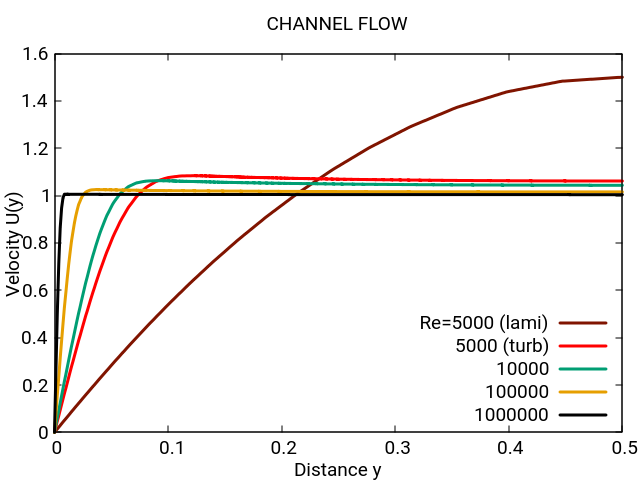}
\includegraphics[width=0.48\textwidth]{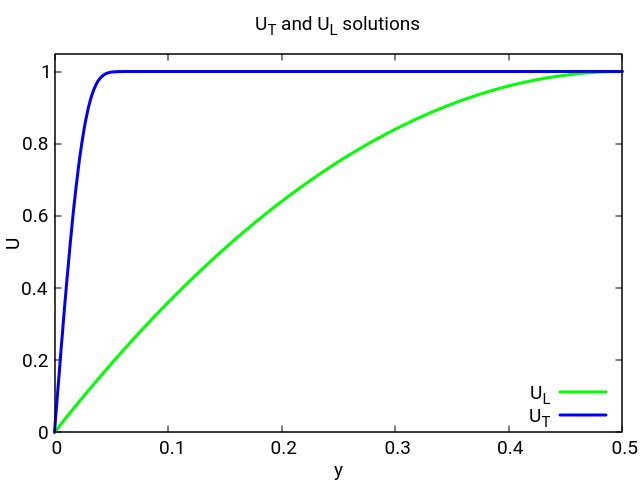}
\hspace{4cm} (a) \hspace{8cm}(b)
\end{center}
\vspace{-7mm}
\caption{\label{fig:utul}(a) Simulation of  2D channel flow where the turbulent solution has been discovered, Fedoseyev and Alexeev (2012) \cite{Fedoseyev_2010}: horizontal velocity U(x), for $Re = 5 \cdot 10^3$ both laminar and turbulent solutions are shown (the brown and red lines), $10^4, 10^5$ and $10^6$ (each solution has the same total flux).
(b) Example of solutions for laminar and turbulent flows, $U_L$,
 Eq.(\ref{poiseille}) (parabolic solution, green line), and $U_T$, Eq.(\ref{eq:UT}) (super exponential solution, blue line). }
\end{figure}

The equations \eqref{cont},\eqref{mom1},\eqref{mom2} also have a second solution, that was initially discovered
numerically for a flow in 2D channel by Fedoseyev and Alexeev (2012) \cite{Fedoseyev_2010}, Figure \ref{fig:utul}(a): both turbulent and laminar numerical solutions are shown for $Re=5000$.

In this paper the second solution is obtained analytically, it is called as a turbulent solution, and it is a super exponential function:
 
\begin{eqnarray} \label{eq:UT}
U_{T}=V_{0}\left(1-e^{1-e^{y/\delta}}\right).
\end{eqnarray}

Figure \ref{fig:utul}(b) shows example of both  $U_L$ (Eq.(\ref{poiseille}), parabolic, green line) and $U_T$ (Eq.(\ref{eq:UT}), super exponential, blue line) solutions for laminar and turbulent flows respectively.

In Section \ref{sec:AGHE} it is explained how to obtain this approximate analytical solution (second solution). 
The construction of general solution of GHE is proposed in Section \ref{sec:general}, and a comparison of this analytical solution with the experimental data is provided in Section \ref{sec:exp}.
%
%
\section{Analytical Solution of GHE\label{sec:AGHE}}

%
%
\subsection{Exact Solution of GHE\label{seq:excat}}
A general solution can be constructed from a particular solution of a problem with non-zero
right hand side (pressure gradient) and general solution with zero
pressure gradient.
Below, it is  described how to find a second solution of GHE and then obtain a general solution.

\noindent
We consider stationary flow regime
\vspace{-2mm}

\begin{eqnarray}
u\,u_{x}+v\,u_{y}+p_{x}&=&Re^{-1}(u_{xx}+u_{yy})\label{mom21}\\
u\,v_{x}+v\,v_{y}+p_{y}&=&Re^{-1}(v_{xx}+v_{yy})\label{mom22}\\
u_{x}+v_{y}&=&\tau(p_{xx}+p_{yy})\label{cont2}
\end{eqnarray}

Let us designate $\alpha$ =$Re^{-1}$ and assume $u_{x}=v_{x}=0,p_{x}=$cont.

Then Eq.(\ref{mom22}, \ref{cont2}) become
\begin{eqnarray}
p_{x}&=&\alpha v_{yy} - v u_y\label{eq:uy}\\
\tau p_{yy}&=&v_{y}\label{eq:p}
\end{eqnarray}

Integrating Eq.(\ref{eq:p}) one obtain
\begin{eqnarray}
\tau p_{y}&=&v +C_1 \label{eq:p2}
\end{eqnarray}

As $p_y=0,v=0$ at the wall, then $C_1=0$. 
Substitute that in Eq.(\ref{eq:uy}) we get
\begin{eqnarray}
\alpha v_{yy}-v v_y - v /\tau = 0\label{eq:uyy}
\end{eqnarray}

\begin{figure}[ht!]
\begin{center}
\includegraphics[width=0.32\textwidth]{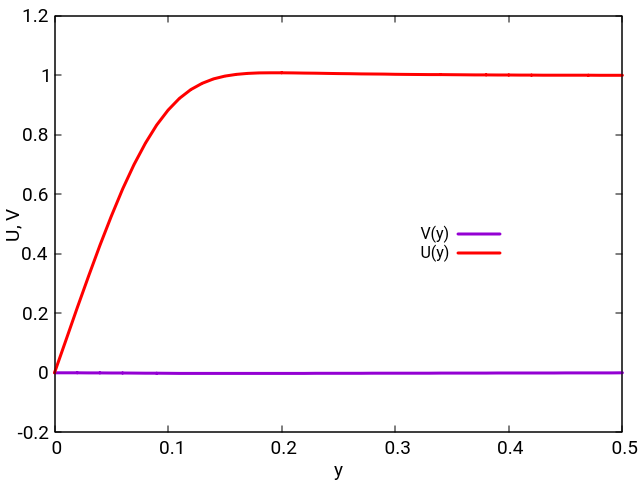}
\includegraphics[width=0.32\textwidth]{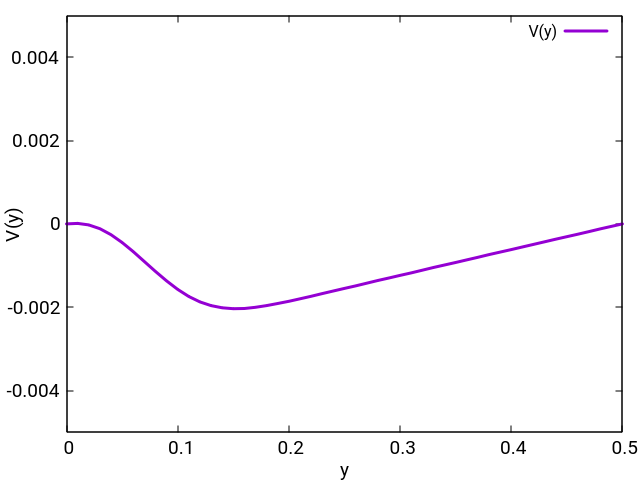}
\end{center}
\vspace{-7mm}
\caption{\label{fig:v}Example of the exact solution $v(y)$ for some $C_1$, $C_2$ of Eq.(\ref{eq:solution}), and $u(y)$  from Eq.(\ref{mom21}): (a) $u(y), v(y)$; (b) $v(y)$ details are shown.}
\end{figure}

The analytical solution of Eq.(\ref{eq:uyy}) is available (e.g. \cite{Polyanin_2002}, Sec.2.2.3-2, Eq.2, and Sec.1.3.1-2 Eq.1, or WolframAlpha.com) in implicit form as Eq. (\ref{eq:solution}),
\begin{equation}\label{eq:solution}
y+C_2 = - \int_{1}^{v(y)} \tau / (W ( -e^{-\tau(\xi^2+2\alpha \tau C_1)/(2\alpha)-1} ) +1) d \xi
\end{equation}

\noindent
where $W(z)=W[k,z]$ is Lambert W function (or the Product Log function), the solution for $w$ in $z= w \exp(w)$. Constants $C_1$ and $C_2$ are chosen to satisfy boundary conditions. There is the $k-th$ solution of the Eq.(\ref{eq:uyy}) corresponding to $W[k,z]$ in Eq.(\ref{eq:solution}). 
There are multiple solutions of $v(y)$, and then multiple solutions $u(y)$ can be found by substituting $v(y)$ into Eq.(\ref{mom21}). The pressure can be found using Eq.(\ref{eq:p2}).
This can not be done analytically to obtain the explicit formulas for $u(y)$ and $p(y)$. 
The example of solutions for $v(y)$ and $u(y)$ are shown in Figure \ref{fig:v}.

In the  Section \ref{sec:approx} we obtain explicit analytical expressions for the approximate solution.

%
%
\subsection{Approximate Solution of GHE\label{sec:approx}}

\vspace{-1mm}
\noindent
Let us designate $\Psi=v_{y}$, $\alpha$ =$Re^{-1}$ and assume $u_{x}=v_{x}=0,p_{x}=$cont,
and neglect small $v\,v_{y}$ term. Then

\begin{eqnarray}
p_{x}&=&\alpha u_{yy}-vu_{y} \label{mom31}\\
p_{y}&=&\alpha\Psi_{y} \label{mom32}\\
\tau p_{yy}&=&\Psi \label{cont3}
\end{eqnarray}

\noindent
Substituting Eq.(\ref{mom32}) into Eq.(\ref{cont3}) obtain

$\alpha\tau\Psi{}_{yy}=\Psi$, or

$\alpha\tau\Psi_{yy}-\Psi=0$.

The solution is $\Psi=Ae^{\pm\frac{y}{\sqrt{\alpha\tau}}}$, and recalling what is $\Psi$:

$v_{y}=Ae^{\pm\frac{y}{\sqrt{\alpha\tau}}}$, integrating that one obtains:

$v=A_{1}e^{\pm\frac{y}{\sqrt{\alpha\tau}}}+B_{1}$, then applying boundary condition:

$v=0$ at y=0, obtain the solution $v$:
\begin{eqnarray} \label{eq:v}
v=A(1-e^{\pm\frac{y}{\sqrt{\alpha\tau}}}) \label{vex},
\end{eqnarray}

Here 
\begin{eqnarray*}
\alpha\tau=\tau Re^{-1}=\frac{\tau^{*}V}{L}\cdot\frac{\nu}{VL}=\frac{\tau^{*}\nu}{L^{2}}=\delta^{2},
\end{eqnarray*}
or 
\begin{eqnarray}
\delta = \frac{\sqrt{\tau^{*}\nu}}{L}
\end{eqnarray}
%
%
\subsection{Case of the Positive Sign in the Exponent}

In this case the solution for $v=A (1-e^{{y}/{\delta}})$ grows exponentially, still for $y \le L/2$ it is limited. So we try to get the solution $u$ using obtained solution for $v$. Let us recall Eq.(\ref{mom31}) and put the solution for $v$ there, we have:

$\alpha u_{yy}-A(1-e^{ {y}/{\delta}})u_{y}-p_{x}=0,$ where $p_{x}=0$,
as we search for a solution of a homogeneous equation

\begin{equation} \label{eq:u}
\alpha u_{yy}-A(1-e^{ {y}/{\delta}})u_{y} = 0,
\end{equation}
To integrate once in $y$ we rewrite Eq.(\ref{eq:u}) as:

\begin{equation*} 
\frac{u_{yy}}{u_{y}}=\frac{A}{\alpha} (1-e^{ {y}/{\delta}}),
\end{equation*}
and obtain integral as
\begin{equation*} 
\ln u_{y}=\frac{A}{\alpha}(y-\delta e^{ {y}/{\delta}})+C_1
\end{equation*}
or
\begin{equation*} 
\ln u_{y}=\frac{A \delta}{\alpha}( {y}/{\delta}- e^{ {y}/{\delta}})+C_1
\end{equation*}
or
\begin{equation} \label{eq:uy2}
u_{y}=\exp(\frac{A \delta}{\alpha}( {y}/{\delta}- e^{ {y}/{\delta}})+C_1)
\end{equation}

\noindent
The explicit expression for the solution of $u(y)$ (integral of $u_y$, Eq.(\ref{eq:uy2})) is available if

\begin{equation} 
\label{eq:A}
A = \frac{\alpha}{\delta}.
\end{equation}

Since $A$ is an arbitrary constant, we may choose $A$ as in (\ref{eq:A}). With this choice, the general solution is

\begin{equation*} \label{eq:u(y)}
u(y) = -\delta e^{C_1} e^{-e^{ {y}/{\delta}}} + C_2.
\end{equation*}

\noindent
If $u(0)=0$, and $u(\infty)=V_{0}$, then the constants $C_1$ and $C_2$ are 
\begin{eqnarray*}
C_2 = V_0,
\end{eqnarray*}
and
\begin{eqnarray*}
\delta e^{C_1} = V_0 e^{1}.
\end{eqnarray*}

We get the solution for $u(y)$ that we call a turbulent solution $U_T$  as:
\begin{align*}
U_{T}=V_{0}\left(1-e^{1-e^{y/\delta}}\right),
\end{align*}
\noindent
that is limited, $u(0)=0$,  $u(L/2) \approxeq V_{0}$, and is acceptable.

%
%
\subsection{Case of the Negative Sign in the Exponent}

Again recall Eq.(\ref{mom31}) and putting the solution for $v$ there, we have:

$\alpha u_{yy}+A(1-e^{-\frac{y}{\delta}})u_{y}-p_{x}=0,$ where $p_{x}=0$,
as we seek a solution of a homogeneous equation.

$\frac{\alpha u_{yy}}{u_{y}}=A(1-e^{-\frac{y}{\delta}}),$

\noindent
Solution for $u_{y}$ :

$\ln u_{y}=Ay+A\delta e^{-\frac{y}{\delta}}+C$

or

$\ln u_{y}=A\delta(\frac{y}{\delta}+e^{-\frac{y}{\delta}})+C$, and

$u_{y}=\exp(A\delta(\frac{y}{\delta}+e^{-\frac{y}{\delta}}))=D\exp(\frac{y}{\delta}-e^{\frac{y}{\delta}})$.

$u=\int\exp(A\delta(\frac{y}{\delta}+\exp(-\frac{y}{\delta}))dx=A_{1}\left[\exp(\frac{y}{\delta}+\exp(\frac{y}{\delta}))-Ei(\exp(-\frac{y}{\delta}))\right]+A_{2}.$

\noindent
As $u(0)=0$, then $A_{2}=0$ and solution $u$ is the following:
\begin{eqnarray}
u=A_{1}\left[e^{\frac{y}{\delta}+e^{\frac{y}{\delta}}}-Ei(e^{-\frac{y}{\delta}})\right], 
\end{eqnarray}
where Ei(x) is the integral exponential function.
This solution for $u$ grows indefinitely, so it is not physical, and will be
disregarded.
%
%
\section{General Solution for the Turbulent Flow\label{sec:general}}

The general solution is proposed as a linear superposition of two solutions, laminar and turbulent:

\begin{eqnarray}\label{eq:GHE1}
U_{GHE}=\gamma U_{T}+(1-\gamma)U_{L}, 
\end{eqnarray}

where the coefficients $\gamma$ and $(1-\gamma)$ are introduced, to get $V_0$ at the center line. The first term containing $U_T$ grows superexponentially in the boundary layer, while the second term containing $U_L$ is nearly zero. Outside the boundary layer, the first term is constant, and the second term starts to grow (see  Figure \ref{fig:exp3a}).

We get 

\begin{eqnarray}\label{eq:GHE2}
U_{GHE}=V_{0}\left[\gamma\left(1-e^{1-e^{y/\delta}}\right)+(1-\gamma)4y(L-y)/L^{2}\right]. 
\end{eqnarray}

The value of $\gamma$ can be calculated from the Eq.(\ref{mom31}) at some point, better for small $y$, for example, $y=\delta$.

Substituting the solution $U_{GHE}$ into Eq.(\ref{mom31}) we get

\begin{eqnarray}
 \alpha {U_{GHE},}_{yy}(\delta) - v(\delta) {U_{GHE},}_y (\delta)  = p_x
\end{eqnarray}

or, dropping the argument $\delta$ , and substituting the expression of $U_{GHE1}$ from Eq.(\ref{eq:GHE1}):

\begin{eqnarray*}
 \alpha (\gamma {U_{T},}_{yy} + (1-\gamma ){U_L,}_{yy})- v (\gamma {U_{T},}_y +(1-\gamma ) {U_L,}_y) = p_x,
\end{eqnarray*}
or
\begin{eqnarray*}
 \gamma (\alpha  ({U_{T},}_{yy} - {U_L,}_{yy}) -v ({U_{T},}_y - {U_L,}_y)) + \alpha {U_L,}_{yy} - v   {U_L,}_y  = p_x,
\end{eqnarray*}
and
\begin{eqnarray}\label{eq:gamma}
\gamma = (p_x - \alpha {U_L,}_{yy} + v  {U_L,}_y) / (\alpha  ({U_{T},}_{yy} - {U_L,}_{yy})
-v ({U_{T},}_y - {U_L,}_y))
\end{eqnarray}
evaluated at $y = \delta $. Here ${U_T,}_{yy}$ and ${U_L,}_{yy}$ are the 2nd partial derivatives of ${U_T}$ and $U_L$ in $y$, ${U_T,}_y$ and ${U_L,}_{y}$ are the first partial derivatives of ${U_T}$ and $U_L$ in $y$, $v(y)$ is calculated using the Eq.(\ref{vex}) with $A$ from Eq.(\ref{eq:A}), $p_x=-8 V_0 / (Re L^2)$. 

The parameter $\delta$ and the value of $\gamma$ in the comparisons with the experiments  below will be chosen to fit the
experimental data, as we do not have the material property
$\tau^*$ for particular liquids used in the experiments, as $\delta = {\sqrt{\tau^{*}\nu}}/{L}$. The value of $\gamma$ can be calculated from the  parameters of experiments using Eq.(\ref{eq:gamma}) if the values of $V_{0}=\frac{1}{2}Re\ p_{x}$ and $L$ are known.

Each of the two solutions has different similarity  parameters. The laminar parabolic solution depends on the Re number,
while the turbulent solution depends on the parameter $\delta= {\sqrt{\tau^{*}\nu}}/{L}.$

The parameter $\tau^{*}$ (dimension of time) is a mean
time between particle collisions.
For gases this  mean time between collisions can be calculated using hard-sphere model of particles at pressure p, and viscosity $\mu$ as $\tau^{*}=\Pi \mu / p$, where $\Pi$ is a constant, $\Pi  = 0.786 $ \cite{Cercignani_1975}. For liquids this is more complicated (see \cite{Alexeev_2004}, p.324 discussion about Frenkel work \cite{Frenkel_1945}), there is no explicit formula, and the experimental measurements are needed.
It can be found by fitting the experimental velocity data (as will be shown in the next Sections) finding $\delta$. Then $\tau^{*}$ can be found as  $\tau^{*} = \delta^2 L^2/ \nu$.
%
%
\section{Comparison of Analytical Solution with Experiments \label{sec:exp}}
%
%
\subsection{Pipe Flow Experiment by Van Doorne at Re=720}
\begin{figure}[ht!]
\begin{center}
\includegraphics[width=0.60\textwidth]{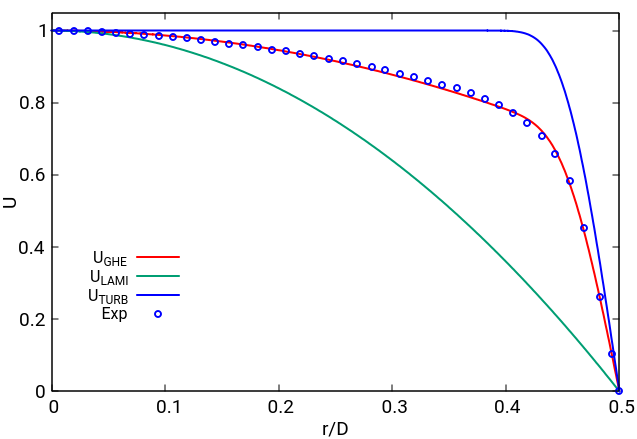}
\end{center}
\vspace{-7mm}
\caption{\label{fig:exp1}
 A comparison of experimental data for  horizontal velocity by Van Doorne(2007) \cite{Doorne_2007}, water
at Re=720, with the GHE model. The GHE solution which is the linear
superposition of the turbulent and lamina profiles ($U_{GHE}$ on the
plot) fits the experimental velocity profile.}
\end{figure}

Here a comparison will be provided for the approximate GHE and other analytical solutions
with the pipe flow data Van Doorne {2004, 2007} \cite{Doorne_2007, Doorne_2004} (Prof. Bruno Eckhardt group). The analytical solution was obtained for 2D flow in Cartesian coordinates, but still we try it for pipes (cylindrical coordinates) assuming that in the vicinity of the wall it will work.
Figure \ref{fig:exp1} shows the experimental data digitized from \cite{Doorne_2007}, and a number
of different plots from (a) laminar (parabolic) flow profile (green line); (b) turbulent solution (blue line) and
and (c) GHE solution. The figure demonstrates that neither
the laminar nor turbulent solution fit the data, but the superposition $U_{GHE}$ provides
an excellent fit to the experimental mean velocity profile for $\gamma$=0.66
and $\delta$=0.047, and $\delta$ is assumed to be a Kolmogorov length scale. In \cite{Fedoseyev_2010} the GHE solution compared well with the experiment for the driven cavity flow \cite{Koseff_1984}, and the dimensional value of $\delta = 0.58mm$ was a good approximation to the experimentally observed {"}Kolmogorov microscale{"} $\delta_{exp} = 0.5mm$ ( \cite{Koseff_1984}, p. 398).

The next Figure \ref{fig:exp2} is a comparison of the values of experimental turbulent intensity for horizontal velocity rms(u\textquoteright )
with the corresponding analytical turbulent intensity. As the solution of the unsteady equations is not provided (not possible analytically), it is suggested that a turbulence is the oscillation between different stationary solutions, two solutions in our case, the laminar (parabolic) velocity profile $U_{L}$ and the turbulent one given by $U_{T}$. Therefore, the rms(u\textquoteright ) at every point is
\begin{equation*}
rms(u^{'}) = \sqrt{\frac{\sum{(U_m-u(t_n))^2}}{N}}
\end{equation*}
\noindent
where $U_m(y)$ is a mean velocity, $u(y,t_n)$ is a velocity at time $t_n$, and $N$ is the number of measurements. 

The analytical rms(u\textquoteright ) was presented as a difference between two solutions $U_{T}$ and $U_{L}$:  
\begin{eqnarray}
rms(u\text{\textquoteright}) =   {\mid \theta \cdot U_{T} - \beta \cdot U_{L}\mid }  \label{rms}
\end{eqnarray}
\noindent
and shown in Figure \ref{fig:exp2}. As one can see, the data and analytical curve are in concordance. The parameters of the analytical curve that define the difference of $U_{T}$ and $U_{L}$ are the following: 
$\theta =3.552,  \beta =2.736,  \delta = 0.047$.

\begin{figure}[ht!]
\begin{center}
\includegraphics[scale=0.60]{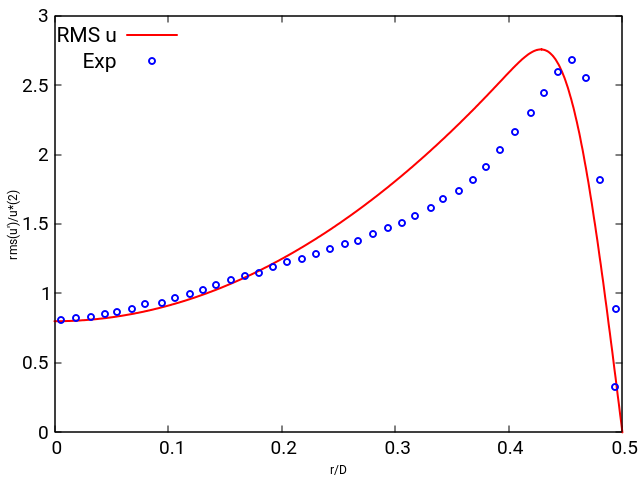}
\end{center}
\vspace{-7mm}
\caption{\label{fig:exp2}Comparison of experimental rms(u') data \cite{Doorne_2007} (blue circles) turbulent intensities
and analytical solution from GHE (red curve). The comparison is quite satisfactory.}
\end{figure}
%
%
\subsection{Experiments by Wei and Willmarth}

\begin{figure}[ht!]
\begin{center}
\includegraphics[width=0.75\textwidth]{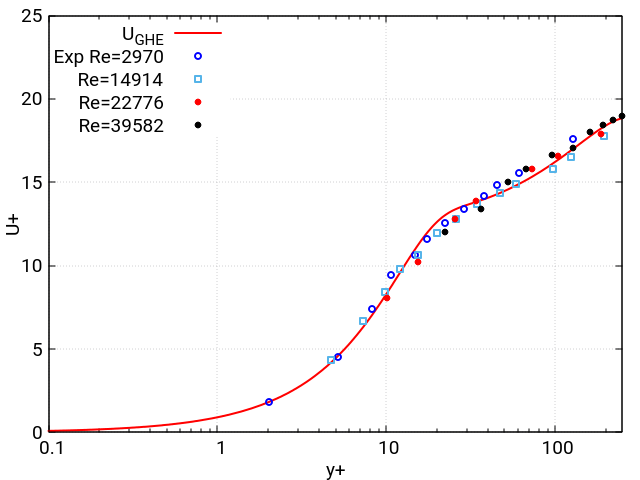}
\end{center}
\vspace{-7mm}
\caption{\label{fig:exp3}Mean velocity profiles in turbulent boundary layer from experiment \cite{Wei_1989} (distilled water), non-dimensionalized on inner variables,
for the four Reynolds numbers examined. Re = 2970(circles); Re = 14914
(squares); Re = 22776 (red dots); Re = 39582 (black dots), distilled
water. The data for different Re numbers merge well in coordinates $u^{+},y^{+}$. The red line is an approximate GHE solution.}
\end{figure}

For the experiments with a turbulent boundary layer, the velocity $U^+$ plots versus $y^+$ does
not depend on Re number, see e.g. data from Wei and Willmarth (1989) \cite{Wei_1989}, Figure \ref{fig:exp3}. The
parameter $y^{+}=\frac{yu_{\tau}}{\nu}$ where $u_{\tau}$ is so called
friction velocity, y is the absolute distance from the wall, and $\nu$
is a kinematic viscosity. One can interpret $y^+$ as a local Reynolds
number. The velocity scale $u_{\tau}$ is defined as $u_{\tau}=\sqrt{\frac{\tau_{w}}{\rho}}$,
where wall shear stress $\tau_w$, $\tau_{w}=\rho\nu\frac{dU}{dy}$
at y=0, and the dimensionless velocity is given by $u^{+}=\frac{u}{u_{\tau}}$. 

The analytical solution $U_{GHE}$ is presented in Figure \ref{fig:exp3} as red line, and is in a good concordance with data (parameters are $\delta=0.052, \gamma=0.65$).

Typically, Figure \ref{fig:exp3} and Figure \ref{fig:exp3a}, the inner boundary layer (BL) region (viscous sublayer) is in the range $0 < y^+ < 5$, where $u^+ = y^+$ (blue line in Figure \ref{fig:exp3a}). The near-middle
(buffer) BL region is in the range $5 < y^+ < 30$, a strictly nonlinear
region. The far-middle (inner) BL region is in the range $30 < y^+ < 200$. Velocity profile in this region can be expressed as the logarithmic
\textquotedblleft law-of-the-wall\textquotedblright{} von Karman law
\begin{eqnarray}
U^+ = 1/k ~log~ y^+ + B, 
\end{eqnarray}
k = 0.41, B = 5.2 (green line in Figure \ref{fig:exp3a}). 

The outer (non-linear, essentially inviscid)) region starts at $y^+ > 200$ and continues to the center line. 

The analytical solution $U_{GHE}$ fits well to the experimental data in all the regions  (Figure \ref{fig:exp3a}, red line).

\noindent
\begin{figure}[ht!]
\includegraphics[width=0.48\textwidth]{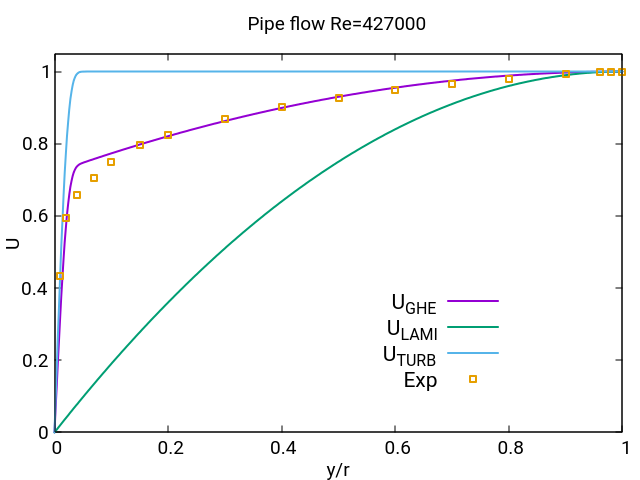}
\includegraphics[width=0.48\textwidth]{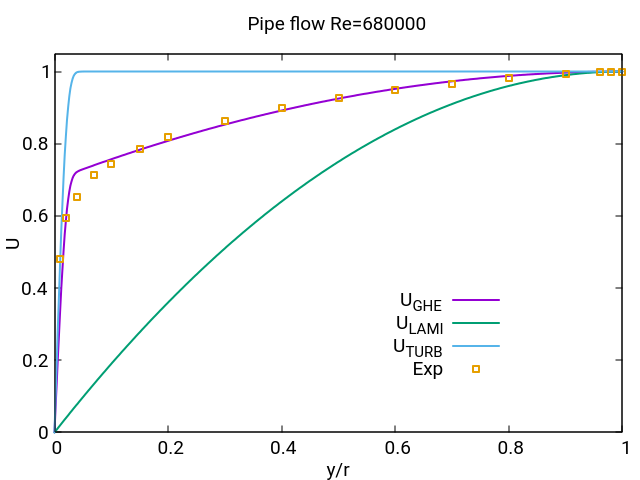}
\hspace{4cm} (a) \hspace{8cm}(b)
\caption{\label{fig:exp4} Experimental data for mean velocity profiles by  horizontal velocity Nikuradse \cite{Nikuradse_1933} for Re=427,000 (a) and Re=970,000 (b) compared to laminar flow profile (green), Turbulent flow profile (blue) and analytical solution GHE flow profile (purple). Approximate analytical solution
$U_{GHE}$ (purple curve) compares well with the experimental data (circles).}
\end{figure}
%
%
\subsection{Experiments by Nikuradse for High Reynolds Numbers}

The solution superposition principle works well for higher values of Reynolds number, up to nearly $10^6$, a range of nearly 3 orders of magnitude. Nikuradse (1932, 1933) (Prandtl group) experiments have been done for Re from 27,000 to Re=970000 \cite{Nikuradse_1933,Nikuradse_1932}. We present a comparison for Re=427,000 and Re=970,000 in Figure \ref{fig:exp4}. The approximate analytical solution
$U_{GHE}$ (pink curve) compares well with the experimental data. The parameters for the analytical solution are the same for all the cases in \cite{Nikuradse_1933},  $\gamma=0.71, \delta=0.010$.
%
%
\section{Discussion \label{sec:discussion}}

\begin{figure}[ht!]
\begin{center}
\includegraphics[width=0.70\textwidth]{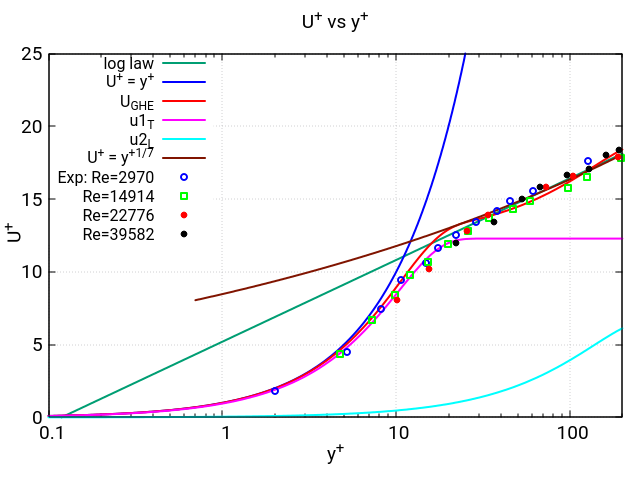}
\end{center}
\vspace{-10mm}
\caption{\label{fig:law}\label{fig:exp3a}
Velocity profiles in turbulent boundary layer: analytical solution $U_{GHE}$ (red line) and its constituents, the turbulent solution $U_{1T}$ and the laminar solution $U_{2L}$; log law by von Karman $U^+$ = 1/k log $y^+$ + B, k = 0.41, B = 5.2 (green line), linear law $U^+ = y^+$ (blue line) and empirical power law $U^+ = {{y^+}~^{1/7}}$ (brown), and experimental data (dots) by Wei and Willmarth (1989)\cite{Wei_1989}. }
\end{figure}

Let us consider the turbulent boundary layer regions, the linear and  the logarithmic \textquotedblleft law-of-the-wall \textquotedblright  von Karman
law parts. Let us represent the turbulent boundary layer profile, Figure \ref{fig:exp3a},  where  the components of GHE with the appropriate coefficients from Eq. (\ref{eq:GHE1}) are plotted: the turbulent solution $U_{1T}$ and the laminar solution $U_{2L}$, were added. The plot  also has the log law by von Karman (the coefficients have been calculated by Nikuradse using his data \cite{Nikuradse_1933}), linear law and  empirical 1/7 power law, that was proposed first by Nikuradse \cite{Nikuradse_1933}. 

The linear law is in the range $0 < y^+ < 5$, where the parabolic profile $U_{2L}$ (laminar solution) is very small, and the turbulent solution for small $y/\delta$ becomes 
\begin{eqnarray*}
U_{T}&=& V_{0}\gamma\left(1-e^{1-e^{y/\delta}}\right) =
V_{0}\gamma\left(1-e^{1-(1+y/\delta)}\right)  = 
V_{0}\gamma\left(1-e^{-y/\delta)}\right) = V_{0}\gamma\left(1-(1-y/\delta)\right) 
\end{eqnarray*}
or
\begin{eqnarray}
U_{T} &=& V_{0}\gamma y/\delta,
\end{eqnarray}
that is a linear law.

\noindent


In the range of $30 < y^+ < 200$, the solution $U_{1T}$ becomes nearly constant $\approx$ 12.5  (pink line in Figure~\ref{fig:law}), and the GHE solution changes due to the growth of the laminar solution $U_{2L}$ (cyan color line in Figure~\ref{fig:law}). That $u^+$ = (12.5 + log $U_{1L}$), the parabolic solution would fit well into logarithmic \textquotedblleft law-of-the-wall\textquotedblright  (von Karman
law). 
The empirical 1/7 power law, that was first proposed by Nikuradse, also fit  well into the \textquotedblleft law-of-the-wall \textquotedblright . 

The parameter $\tau^{*}$, a material property, if not known from the experiment for a given fluid, 
can be found by fitting the experimental velocity data (as we did in the previous Sections). In this case we will obtain the value of $\delta$ which will be used to calculate $\tau^{*}$:  $\tau^{*} = \delta^2 L^2/ \nu$.
%
%
\section*{Conclusions}

The exact and approximate analytical solution of Generalized Hydrodynamic Equations (GHE) have been obtained for channel flow (that are valid for the pipe flow if $\delta \ll 1$), that constructed from two solutions for laminar and turbulent flows, presented as a linear superposition of these solutions. 

The GHE solution for turbulent flows depends on two parameters: the Reynolds number $Re$, and $Al=\frac{\sqrt{\tau^{*}\nu}}{L} = \frac{\delta}{L}$, the Alexeev number (author of GHE and GBE). 
In Figure~\ref{fig:law}, the data for different $Re$ numbers fall to a single law in ($U^+ , y^+$) coordinates for the same kind of fluid. This data has the same Alexeev number $Al$ if the same experimental dimensions hold. 
Changing the $Al$ number (different fluid or dimensions) will  produce different results. 

The time parameter $\tau^{*}$ is a material property. For gases this is a mean time between collisions, and can be calculated for hard-sphere model of particles at pressure p, and viscosity $\mu$ as $\tau^{*}=\Pi \mu / p$, where $\Pi$ is a constant, $\Pi  = 0.786 $. For liquids this is more complicated (see \cite{Alexeev_2004}, p.324 discussion about Frenkel work \cite{Frenkel_1945}), there is no explicit formula, and the experimental measurements are needed.

The approximate analytical solution for the GHE model has been  compared with the experimental data for turbulent flows. The linear superposition of two solutions (turbulent and laminar) provides good agreement with the experimental axial mean velocity data in a wide range of Reynolds number from Re=720 to Re = 960,000. 

The experimental data for turbulent intensities for axial velocity are in concordance with the analytical equation (weighted difference of the turbulent and laminar solutions). Apparently the turbulent oscillations are the oscillations between two solutions (two in considered case): the turbulent solution $U_T$ and the laminar solution $U_L$.  GHE intrinsically include a turbulence model.

Moreover, the obtained analytical solution was able to capture correct velocity behavior across the whole turbulent boundary layer (BL) and into the external flow: from the inner viscous sublayer to the outer layer of the turbulent BL. It also compares well with the experiments by Wei and Willmarth (1989). 

The analytical solution  presents the explicit verifiable proof that the Alexeev generalized hydrodynamic theory (GHE and GBE) is in good agreement with experiments for turbulent flows.


\vspace{5mm}
\noindent
{\bf Availability of data}

\noindent
The data that support the findings of this study are available from the corresponding author upon reasonable request.
%
%
\section*{Acknowledgments}
Author would like to extend the sincere thanks to Dr. Yuri Bozhkov for discussions of the text during manuscript preparation.


\begin{thebibliography}{10}

\bibitem{Alexeev_1994}B. V. Alexeev, The generalized Boltzmann equation,
generalized hydrodynamic equations and their applications. Phil. Trans.
Roy. Soc. London, A. 349 (1994), 417-443. 

\bibitem{Alexeev_2004} Alexeev, B.V.,Generalized Boltzmann Physical Kinetics, Elsevier, 2004.

\bibitem{Allegre_1997} J. Allegre, D. Bisch, J.C. Lengrand, J. of Spacecraft and
Rockets, 34 (6), 724-728 (1997) 

\bibitem{Cercignani_1975}Cercignani C. Theory and Application of the Boltzmann Equation, Scottish Academic Press, Edinburgh and London, 1975.

\bibitem{Doorne_2007}C.W.H. Van Doorne, Jerry Westerweel, 2007, Measurement of laminar, transitional and turbulent pipe flow using Stereoscopic-PIV, February
2007, Experiments in Fluids 42(2), DOI: 10.1007/s00348-006-0235-5.

\bibitem{Doorne_2004} Casimir Willem Hendrik Van Doorne, Stereoscopic PIV on transition in pipe flow, Ph.D. Thesis, TU Delft, Netherlands, 2004.

\bibitem{Fedoseyev_1998} Fedoseyev A.I., Alexeev, B.V.  Mathematical model for viscous incompressible fluid flow using Alexeev equations and comparison with experimental data. In: Dey, S.K., Ziebarth, G., Ferrandiz (Eds.), Proceedings of Advances in Scientific Computing Modelling (Special Proceedings of IMACS{'}98). Alicante, Spain, 1998, 158-163.

\bibitem{Fedoseyev_2000}A. I. Fedoseyev, E. J. Kansa, C. Marin, M. Volz, and A. G. Ostrogorsky, AIAA Paper 2000-0698.

\bibitem{Fedoseyev_2001a}Fedoseyev A., A regularization approach to solving the Navier-Stokes Equations for Problems with Boundary Layer, Comput. Fluid Dynamics J., 9, (2001) 317-324.

\bibitem{Fedoseyev_2001b} A. I. Fedoseyev, E. J. Kansa, C. Marin, and A.G. Ostrogorsky (2001) Japanese Computational Fluid Dynamics Journal 10(3),
325-333.

\bibitem{Fedoseyev_2010} Fedoseyev A.I., Alexeev, B.V.,Simulation of viscous flows with boundary layers within multiscale model using generalized hydrodynamics equations , Procedia Computer Science, 1 (2010)665-672.

\bibitem{Fedoseyev_2012} Fedoseyev A., Alexeev B.V., Generalized hydrodynamic equationsfor viscous flows-simulation versus experimental data,in AMiTaNS-12,
American Institute of Physics AIP CP1487, pp.241-247. 

\bibitem{Fedoseyev_2020} Fedoseyev A., Finite element method stabilization for supersonic flows with flux correction transport method, AIP Conference
Proceedings 2302, 120003 (2020); https://doi.org/10.1063/5.003351 

\bibitem{Fedoseyev_2021} Fedoseyev A., Griaznov V., Simulation of Rarefied Hypersonic Gas Flow and Comparison with Experimental Data, in AMiTaNS-2021, Conf. Proc., AIP CP 2522, 100003, 2021, Ed. M.Todorov. 

\bibitem{Fedoseyev_2022} Fedoseyev A., Griaznov V., Ouazzani J., Simulation of rarefied hypersonic gas flow and comparison with experimental data
II, Proc. AMITANS-2022 Conf., AIP CP 2953, 2023, Ed. M.Todorov. 

\bibitem{Frenkel_1945} Frenkel\textquoteright, Ya.I. . Kineticheskaya Teoriya Zhidkostei, (Kinetic Theory of Liquids).1945 Izd. AN SSSR, Moscow--Leningrad.

\bibitem{Harvey_2001} Harvey J.K., Holden M. S., and Wadhams, T. P., Code Validation Study of Laminar Shock/Boundary Layer and Shock/Shock Interactions in
Hypersonic Flow, Part B: Experimental Measurements," AIAA Paper 2001-1031, January 2001.

\bibitem{Harvey_2003} Holden, M. S., Wadhams, T. P., Candler, G. V., Harvey J. K., Measurements in Regions of Low Density Laminar Shock Wave/Boundary Layer
Interaction in Hypervelocity Flows and Comparison with Navier-Stokes Predictions," AIAA Paper 2003-1131, January 2003.

\bibitem{Kim_1980}J. Kim, S.J. Kline, and J. P. Johnston (1980) ASME J. Fluids Engng. 102, 302-308.

\bibitem{Koseff_1984} R. Koseff, R. L. Street , Trans. ASME/Journal of Fluids Engineering 106 (1984) 390-398. 
\bibitem{Nikuradse_1933} J. Nikuradse, Laws of Flow In Rough Pipes, NASA Technical Memorandum 1292, 1950. Translation of Stromungsgesetze in rauhen
Rohren., VDI-Forschungsheft 361. Beilage zu  Forschung
auf dem Gebiete des Ingenieurwesens{} Ausgabe B Band 4, July/August 1933. 

\bibitem{Nikuradse_1932}J. Nikuradse, Laws of turbulent flow in smooth pipes
(English translation), NASA (1932) TT F-10: p. 359 (1966). Available
from https:// www.Princeton .edu/mae/people/ faculty/smits/homepage/data-1/nikuradse data/Nikuradse German\_1932.pdf 

\bibitem{Polyanin_2002} Polyanin A.D., Zaitsev V.F., Handbook of Exact Solutions for Ordinary Differential Equations, 2002, Chapman Hall, CRC Press Company, Boca Raton, London.

\bibitem{Wei_1989}T. Wei ,W. W. Willmarth, Reynolds-number effects on the structure of a turbulent channel flow, J. Fluid Mech. (1989), vol 204, pp.
57-95.

\end{thebibliography}
\end{document}